# Ring surface waves in thermal nonlinear media

Yaroslav V. Kartashov,[1] Victor A. Vysloukh,[2] and Lluis Torner[1]

[1] *ICFO-Institut de Ciencies Fotoniques, and Universitat Politecnica de Catalunya, Mediterranean Technology Park, 08860 Castelldefels (Barcelona), Spain*
[2] *Departamento de Fisica y Matematicas, Universidad de las Americas – Puebla, Santa Catarina Martir, 72820, Puebla, Mexico*
*Yaroslav.Kartashov@icfo.es*

**Abstract:** We address ring-shaped surface waves supported by defocusing thermal media with circular cross-section. Such waves exist because of the balance between repulsion from the interface and deflection of light from the bulk medium due to defocusing nonlocal nonlinearity. The properties of such surface waves are determined by the geometry of the sample. Nodeless ring surface waves are stable for all values of their winding number, while surface waves with a small number of azimuthal nodes can be metastable.

Nonlocality is an intrinsic feature of many optical materials [1]. Among the strongly nonlocal materials which can be used for the experimental exploration of soliton formation are liquid crystals [2-5] and thermal media [6-8]. Besides the simplest solitons, such materials support several types of stable, higher-order two-dimensional structures [9-13].

Light beams propagating in nonlocal media cause refractive index changes in regions far exceeding the beam width, therefore the beam trajectory can be strongly altered by the presence of interfaces, like in thermal media where the induced refractive index profile is determined by the sample geometry [14,15]. Under proper conditions light may concentrate in the vicinity of the interface, exciting surface waves in both local [16-19] and nonlocal nonlinear materials [20-24]. Two-dimensional surface waves at the interface of focusing thermal medium were observed in Ref [23]. Also, thermal materials with defocusing nonlinearity have been shown to support one-dimensional surface waves [24].

In this paper we reveal that thermal samples of circular cross-section with a defocusing nonlinearity support specific ring-profile surface waves localized near the outer edge of the sample. Such waves form due to the balance between repulsion from the interface and deflection of light from the bulk sample due to a defocusing nonlinearity.

We consider the propagation of laser beam along the $\xi$ axis of a defocusing thermal medium with circular cross-section, described by the equations for the field amplitude $q$ and nonlinear contribution to the refractive index $n$:

$$i\frac{\partial q}{\partial \xi} = -\frac{1}{2}\left(\frac{\partial^2 q}{\partial r^2} + \frac{1}{r}\frac{\partial q}{\partial r} + \frac{1}{r^2}\frac{\partial^2 q}{\partial \phi^2}\right) - qn,$$
$$\frac{\partial^2 n}{\partial r^2} + \frac{1}{r}\frac{\partial n}{\partial r} + \frac{1}{r^2}\frac{\partial^2 n}{\partial \phi^2} = |q|^2. \qquad (1)$$

Here the radial and longitudinal coordinates $r, \xi$ are scaled to the beam radius and the diffraction length, respectively; $\phi$ is the azimuthal angle. Equation (1) was solved with the boundary conditions $q, n|_{r \to R} = 0$ ($R$ is the sample radius; here we set $R = 20$, but results remain valid for other radiuses), corresponding to the situation when light does not penetrate into the regions outside the thermal sample, whose surface is maintained at a fixed temperature. Physically, this can be achieved when the refractive index of the surrounding material, that plays the role of heat sink, is much lower than that of the thermal medium. In this case the beam propagating in the cylindrical rod experiences slight absorption, that causes an increase and a redistribution of temperature in the entire medium due to heat diffusion.

We assume a typical situation where the refractive index decreases in the heated regions. Such a decrease causes light defocusing and thus expulsion of light from the bulk material toward its periphery. In circular samples the repulsion from the interface with the less opti-cally dense surrounding medium results in light localization in the vicinity of interface in the form of bright rings that may feature azimuthal or angular nodes.

Radially symmetric surface waves can be found in the form $q(r,\phi,\xi)=w(r)\exp(im\phi)\times\exp(ib\xi)$, where $m$ is the topological charge, and $b$ is the propagation constant. For such excitations the refractive index is given by

$$n(r,\xi)=\int_0^R G_0(r,\rho)|q(\rho,\xi)|^2\,d\rho, \tag{2}$$

where $G_0(r,\rho)=\rho\ln(\rho/R)$ for $r<\rho$ and $G_0(r,\rho)=\rho\ln(r/R)$ for $r\geq\rho$ is the response function of the medium. For the linear stability analysis we will looking at solutions in the form $q=[w(r)+u(r,\xi)\exp(ik\phi)+v^*(r,\xi)\exp(-ik\phi)]\exp(ib\xi+im\phi)$, where the perturbations components $u,v$ can grow with a complex rate $\delta=\delta_r+i\delta_i$ upon propagation, and $k$ is the azimuthal perturbation index. Substitution into Eq. (1) and linearization yields the eigenvalue problem

$$\begin{aligned} i\delta u &= -\frac{1}{2}\left[\frac{d^2u}{dr^2}+\frac{1}{r}\frac{du}{dr}-\frac{(m+k)^2}{r^2}u\right]-w\Delta n_k - un + bu, \\ i\delta v &= \frac{1}{2}\left[\frac{d^2v}{dr^2}+\frac{1}{r}\frac{dv}{dr}-\frac{(m-k)^2}{r^2}v\right]+w\Delta n_k + vn - bv, \end{aligned} \tag{3}$$

where

$$\Delta n_k = \int_0^R G_k(r,\rho)w(\rho)[u(\rho)+v(\rho)]d\rho \tag{4}$$

is the refractive index perturbation corresponding to the azimuthal index $k$, while $G_k(r,\rho)=-(\rho/2k)[(r/\rho)^k-(r\rho/R^2)^k]$ for $r<\rho$, and $G_k(r,\rho)=-(\rho/2k)[(\rho/r)^k-(r\rho/R^2)^k]$ for $r\geq\rho$.

Nodeless ring-shaped surface waves exist for $b\leq 0$ [as depicted in Fig. 1(a)]. With decreasing $b$, the peak amplitude of the surfaces waves increases, while the separation between the interface and the wave peak intensity decreases. A decrease in $b$ is accompanied by a monotonic growth of the energy flow [as seen in Fig. 1(d)]

$$U = 2\pi\int_0^R rw^2 dr \tag{5}$$

and by an enhanced localization of light near the sample boundary. The degree of localization may be measured by the radial width of the surface wave, which is found to be a monotonically increasing function of $b$ [Fig. 1(e)]. Despite the fact that light is localized near the sample edge, the refractive index exhibits a flat plateau across most of the sample [see Fig. 1(b)], and it is locally increased in the vicinity of the sample edge. Such increase gives rise to surface waves. The decrease in the refractive index in the center of the sample becomes more pronounced with increasing light intensity.

We also found solutions with various number $n_r$ of radial nodal lines [Fig. 1(c)]. For such waves the local intensity maximum located farthest from the edge is always the most pronounced. Despite the presence of radial oscillations in $w$, the refractive index always has a shape similar to those shown in Fig. 1(b). At fixed $b$, the energy flow carried by the wave with larger $n_r$ is smaller [Fig. 1(d)]. Ring surface waves with any $n_r$ can carry screw phase dislocations characterized by a winding number, or topological charge, $m$. Since rings are localized near the outer sample edge and their radii are close to $R$, the profiles of waves with moderate $m$ values are almost identical.

Surface wave localization is achieved due to the balance set between the expulsion from the bulk of the defocusing sample and the repulsion from its edge. Note that in this setting, any potential azimuthal instabilities are suppressed and all nodeless waves are stable for any $m$. The instability suppression occurs because in defocusing media a local increase of the

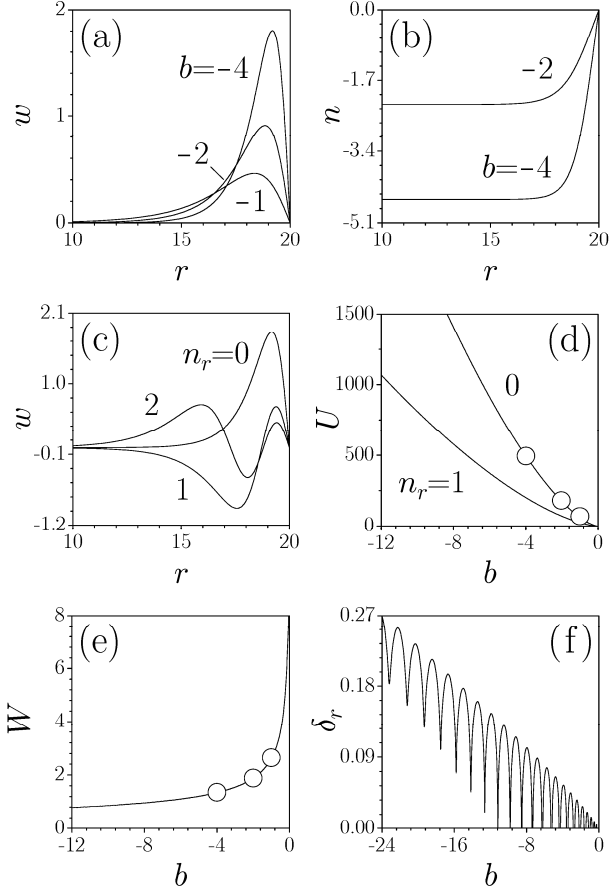

Fig. 1. Profiles (a) and refractive index distributions (b) for nodeless surface waves with $m = 0$, $n_r = 0$ for different $b$ values. (c) Profiles of surface waves with different $n_r$ at $b = -4$, $m = 0$. (d) Energy flow versus $b$ for surface waves with different $n_r$ at $m = 0$. (e) Integral width of surface wave with $n_r = 0$, $m = 0$ versus $b$. Points marked by circles in (d) and (e) correspond to waves shown in (a). (f) Real part of perturbation growth rate corresponding to $k = 1$ for wave with $m = 0$, $n_r = 1$ versus $b$.

light intensity at a given spatial point caused by the azimuthal modulation results in a decrease of the refractive index in the vicinity of this point. Therefore, light tends to deflect from such regions, thus restoring an azimuthally uniform distribution. At the same time, deflection from the bulk sample and repulsion from its edge ensure that light remains radially localized near the sample edge. In our comprehensive numerical experiments, nodeless waves keep their structure for huge distances in the presence of the strong random and regular perturbations

(Fig. 2, top). Note that in all cases shown in Figs. 2 and 3, weak multiplicative azimuthal perturbations $[1 + 0.002\sum_{n=1}^{N}\cos(p\phi)]$ with $N \gg 1$ were added to the stationary field distribu-tions with the aim to trigger any potential azimuthal instabilities. Thus, the input in Fig. 2 that would be radially symmetric without perturbation appears slightly distorted. All higher-order waves with $n_r > 0$ are turn out to be oscillatory unstable. Figure 1(f) shows the dependence of the perturbation growth rate $\delta_r$ on $b$ for the wave with $n_r = 1$. For a fixed azimuthal index $k$ there are regions (close to $b=0$) where $\delta_r$ vanishes, but they overlap with regions where $\delta_r$ is nonzero for other values of $k$. The decay of waves with radial nodes is accompanied by the gradual fusion of their rings (Fig. 2, bottom) that occurs faster at higher energy flows.

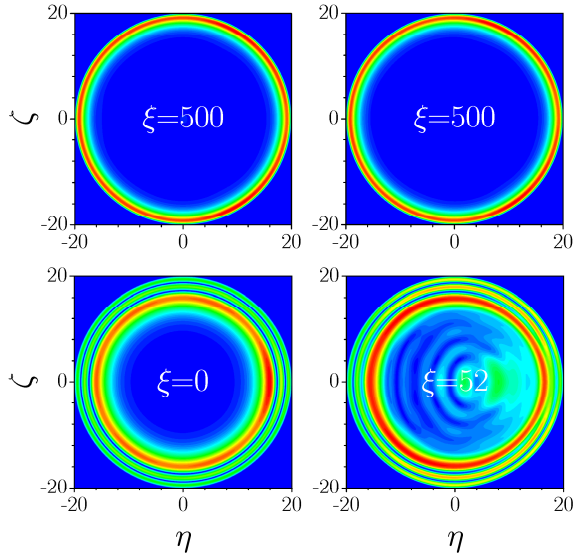

Fig. 2. Propagation of perturbed surface waves with different $n_r$. Field modulus distributions for waves with $n_r = 0$, $m = 0$ (top, left), $n_r = 0$, $m = 3$ (top, right), $n_r = 2$, $m = 0$ (bottom) are shown at different distances. All waves correspond to $b = -4$.

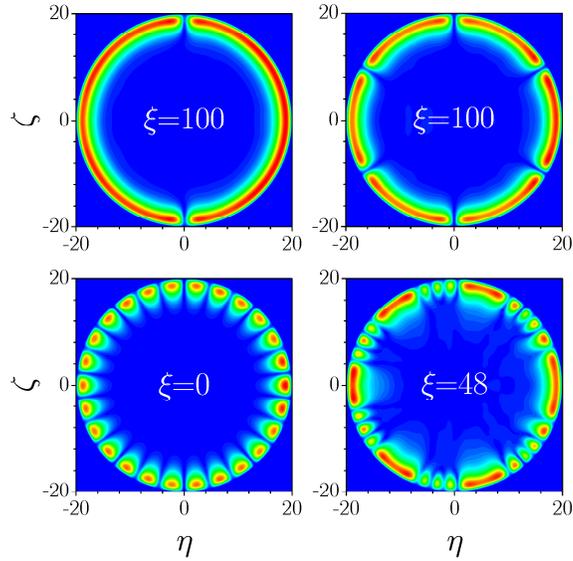

Fig. 3. Propagation of perturbed surface waves with different $n_\phi$. Field modulus distributions for waves with $n_\phi = 2$ (top, left), $n_\phi = 6$ (top, right), and $n_\phi = 26$ (bottom) are shown at different distances. All waves correspond to $b = -2$.

We also found surface waves with azimuthal nodal lines, featuring necklace-like shapes with phase changing by $\pi$ between the bright spots forming such waves. Representative profiles, obtained from Eq. (1) in the form $q(r,\phi,\xi) = w(r,\phi)\exp(ib\xi)$, are shown in Fig. 3. The energy flow of such waves increases with decreasing $b$, while the radial width decreases. The higher the number of nodal lines $n_\phi$ the smaller the energy flow at fixed $b$. Far from the azimuthal nodes, such waves resemble the previously discussed rings, especially for small values of $n_\phi$ (see the top panel in Fig. 3). The refractive index profile exhibits a plateau in the center of the sample, but close to the sample edge it is azimuthally modulated.

Rigorously speaking, all surface waves with $n_\phi > 0$ were found to be unstable. This was confirmed by comprehensive simulations in the presence of regular and noisy perturbations for different $b$ and $n_\phi$ values. In all cases, in our numerical simulations azimuthally modulated solutions self-destroyed after propagation over sufficiently large distances. However, the corresponding perturbation growth rates may be extraordinarily small for waves with small $n_\phi$ values and at moderate energy flows. The decay distances for such waves, when they are strongly perturbed, far exceeds the experimentally available sample lengths (see the top row of Fig. 3 showing a long-distance propagation of perturbed waves with $n_\phi = 2$ and $6$). As expected on intuitive grounds, increasing the number of azimuthal nodes substantially enhances instabilities at fixed $b$. Waves with large $n_\phi$ quickly decay via increasing oscillations of bright spots (as illustrated in the bottom panel of Fig. 3). However, waves with small $n_\phi$ should be readily observed experimentally in short sample lengths.

Summarizing, we have found that the balance between expulsion from the bulk of defocusing thermal media and repulsion from their edge may result in the formation of new types of surface waves, that exhibit ring-like shapes in accordance with the sample geometry.

**Acknowledgements**

This work has been supported in part by the Government of Spain through the Ramon-y-Cajal program and the grant TEC2005-07815, and by CONACYT through the grant 46552.